\documentclass[twocolumn]{aastex63}
\usepackage{amsmath}
\usepackage{graphicx}
\usepackage{graphicx,psfrag,epsfig,subfigure}
\usepackage{url}
\usepackage{indentfirst}
\usepackage{CJK}
\usepackage{txfonts}
\usepackage{paralist}
\usepackage{fancyhdr,lastpage}
\usepackage{color,hyperref}
\usepackage{pslatex}
\usepackage{amssymb}
\usepackage{afterpage}
\usepackage{upgreek}
\usepackage{epstopdf}
\usepackage{longtable}

\makeatletter
\let\jnl@style=\rm
\def\ref@jnl#1{{\jnl@style#1}}
\makeatother

\newcommand*\patchAmsMathEnvironmentForLineno[1]{%
\expandafter\let\csname old#1\expandafter\endcsname\csname #1\endcsname
\expandafter\let\csname oldend#1\expandafter\endcsname\csname end#1\endcsname
\renewenvironment{#1}%
{\linenomath\csname old#1\endcsname}%
{\csname oldend#1\endcsname\endlinenomath}}%
\def\approxgt{\mathrel{\hbox{\rlap{\lower.55ex \hbox {$\sim$}}
        \kern-.3em \raise.4ex \hbox{$>$}}}}
\def\approxlt{\mathrel{\hbox{\rlap{\lower.55ex \hbox {$\sim$}}
        \kern-.3em \raise.4ex \hbox{$<$}}}}

\def\Msun{\hbox{~$M_{\odot}$}}

\def\H0{{\rm ~km~s^{-1}~Mpc^{-1}}}

\def\p9{{Pa$9$}}

\def\six{[\ion{S}{IX}]}

\def\L2-10{L$_{\rm 2-10keV}$}

\def\.25{0.25 keV\thinspace}

\def\d19{D$~\leq~$19~Mpc}

\newcommand{\swift}{\textit{Swift}}
\newcommand{\swiftc}{\swift\ }

\newcommand{\frb}{FRB~121102}
\newcommand{\frbc}{\frb\ }

\begin{document}

\title{A search for hard X-ray bursts occurring simultaneously to fast radio bursts in the repeating FRB 121102}

\correspondingauthor{ Shangyu~Sun and Wenfei~Yu}
\email{sysun@shao.ac.cn, wenfei@shao.ac.cn}

\author[0000-0002-0786-7307]{Shangyu ~Sun}
\affiliation{Key Laboratory for Research in Galaxies and Cosmology, Shanghai Astronomical Observatory, Chinese Academy of Sciences, \\
80 Nandan Road, Shanghai 200030, China.}

\author{Wenfei Yu}
\affiliation{Key Laboratory for Research in Galaxies and Cosmology, Shanghai Astronomical Observatory, Chinese Academy of Sciences, \\
80 Nandan Road, Shanghai 200030, China.}

\author{Yunwei Yu}
\affiliation{Institute of Astrophysics, Central China Normal University, Wuhan 430079, China}

\author{Dongming Mao}
\affiliation{Key Laboratory for Research in Galaxies and Cosmology, Shanghai Astronomical Observatory, Chinese Academy of Sciences, \\
80 Nandan Road, Shanghai 200030, China.}




\begin{abstract}
 
The nature of fast radio bursts (FRBs) is currently unknown.  Repeating FRBs offer better opportunity than non-repeating FRBs since their simultaneous multi-wavelength counterparts might be identified. The magnetar flare model of FRBs is one of the most promising models which predicts high energy emission in addition to radio burst emission. To investigate such a possibility, we have searched for simultaneous and quasi-simultaneous short-term hard X-ray bursts in all the Swift/BAT event mode data which covered the periods when fast radio bursts were reported detections in the repeating FRB~121102, by making use of BAT's arcmin level spacial resolution and wide field-of-view. We did not find any significant hard X-ray bursts which occurred simultaneously to those radio bursts. We also investigated potential short X-ray bursts occurred quasi-simultaneous with those radio bursts (occurrence time differs in the range from hundreds of seconds to thousands of seconds) and concluded that even the best candidates are consistent with background fluctuations. Therefore our investigation concluded that there were no hard X-ray bursts detectable with Swift/BAT which occurred simultaneously or quasi-simultaneously with those fast radio bursts in the repeating FRB~121102.

\end{abstract}


\section{Introduction} \label{intro}

The fast radio burst (FRB) was a recently recognized class of astrophysical transient phenomena \citep{2007Sci...318..777L,2013Sci...341...53T}, with features of short durations (millisecond timescale), high flux densities ($\sim$Jy), and large dispersion measures (DMs) implying cosmological distances. The localization of several FRBs \citep{2017ApJ...834L...7T, 2019Sci...365..565B,2019Natur.572..352R} even certifies their cosmological origin. The \frbc was the first discovered repeating FRBs \citep{2014ApJ...790..101S,2015MNRAS.454..457P,2016ApJ...833..177S,2016Natur.531..202S}. Afterwards, the CHIME/FRB Collaboration \citep{2019Natur.566..230C} discovered a second repeating one \citep[FRB~180814.J0422+73; ][]{2019Natur.566..235C} and soon eight more \citep[][]{2019arXiv190803507T}. Many of these repeating FRBs show more common features in their bursts: complex burst morphology and sub-burst downward frequency drifts \citep{2019Natur.566..235C, 2019ApJ...876L..23H, 2019arXiv190803507T} although these features are also seen in some nonrepeating FRBs \citep[e.g. ][]{2016MNRAS.460L..30C}. The repetition of these sources naturally exclude cataclysmic models for them. However, the physical nature of neither repeating nor nonrepeating FRB sources is clearly understood.

The repetition of the \frbc enabled interferometric localization of it \citep[R.A.: 05$^{\rm h}$31$^{\rm m}$58$^{\rm s}.70$, decl.: +33$^{\circ}$08$^{\prime}$52$^{\prime\prime}.5$; ][]{2017Natur.541...58C} and identification of its host galaxy \citep[$z=0.19273$, $\sim$970~Mpc; ][]{2017ApJ...834L...7T}. The bursts from it show highly variable spectra \citep{2016ApJ...833..177S,2019ApJ...876L..23H}. \cite{2016Natur.531..202S} and \cite{2017Natur.541...58C} reported a dispersion measure (DM) of 558.1$\pm3.3$~pc~cm$^{-3}$. First of all, based on the feature of tremendous energy released in milliseconds and repetition of bursts, its nature is suspected to relate more to a young active neutron star, for example the phenomena of magnetar flare and giant pulse \citep{2010vaoa.conf..129P,2014ApJ...797...70K,2014MNRAS.442L...9L, 2016MNRAS.458L..19C}. Secondly, the high rotation measures (RM) of FRB 121102 \citep[$>10^{5}$~rad~m$^{-2}$; ][]{2018Natur.553..182M} indicate extreme magneto-ionic surroundings of the burst source, and the estimated magnetic field of such large strength can be explained by that of a magnetar \citep[for e.g. SGR~J1745--2900][]{2013Natur.501..391E}. Thirdly, \cite{2017ApJ...834L...7T} identified the host galaxy of \frbc as a compact (diameter $\approxlt$ 4~kpc) dwarf ($\sim 6 \times 10^{7}$~\Msun), which is believed to be the crib of superluminous supernovae or long-duration gamma-ray bursts. They are likely the progenitor of a millisecond magnetar \citep{1998PhRvL..81.4301D,2001ApJ...552L..35Z,2010ApJ...719L.204W,2010ApJ...717..245K,2017ApJ...840...12Y}, which is hence considered to be the origin of \frb. 
Therefore, the young magnetars responsible for FRBs could be formed from some unusual core-collapse supernovae and then they are harbored at the center of a supernova remnant \citep{2017ApJ...841...14M,2018ApJ...868L...4M,2019MNRAS.485.4091M}. For a more general consideration, these young magnetars could also be born from the mergers of double neutron stars \citep{2006Sci...311.1127D,2008MNRAS.385.1455M,2018ApJ...861..114Y} or from the accretion-induced collapse of white dwarfs \citep{1990ApJ...356L..51C, 1991ApJ...367L..19N}. This possibility at least has been somewhat supported by the observations of the non-repeating FRBs \citep{2018ApJ...858...89C,2019ApJ...886..110M}. In this case, the young magnetars cannot be associated with a supernovae remnant, but can still be surrounded by a pulsar wind bubble \citep{2017ApJ...838L...7D}. Therefore, no matter which channel the magnetars originate from, a persistent radio counterpart emission can be always generated by the interactions between the ejecta, pulsar wind, and the surrounding medium. By confronting with the flux of the persistent radio emission and the decreasing DM of \frbc \citep[i.e., 10\% decreasing in seven months][]{2018Natur.553..182M}, the age of the neutron star can be tightly constrained to be about one hundred years old \citep{2017ApJ...839L..20C,2017ApJ...841...14M}. In view of its highly relevant to a young magnetar, it is very natural to expect FRB 121102 could be accompanied by some high-energy emissions.

The \swift/BAT can cover the entire sky as efficiently as 80--90$\%$~per day \citep{2013ApJS..209...14K} and offers a time resolution of $10^{-4}$ s for trigger event mode data. It is good for detecting or monitor short-and-bright X-ray transient sources, and therefore meets the requirement for addressing the aforementioned scientific questions. 
A blind search for hard X-ray bursts in the BAT data of one-year time (2016 October 1 to 2017 September 30) had been conducted \citep{2019ApJ...885...55S}. Continuing the previous efforts, we searched for X-ray bursts in all the BAT trigger event mode data that were simultaneous to all the radio bursts from \frbc that had been reported in the  literature. In addition to the search for simultaneous X-ray bursts, we also performed a search for possible bursts which did not occur simultaneously in these data. We present the method and our results achieved in our search for burst signals in the direction of FRB 121102 with the \swift/BAT data in Sect.~\ref{obser}. Then we discuss and conclude in Sect.~\ref{dc}.

\section{Observations and analysis} \label{obser}

\begin{table*}
\caption{Radio burst detected from \frbc since 2012-11-02.}\label{T:bursts}
\centering
\begin{tabular}{cccccc}
\hline 
  \textbf{reference}  & \textbf{$N_{\rm rb}$ $^{a}$}        & \multicolumn{2}{c}{\textbf{dates of bursts}}  & \textbf{radio}                         & \textbf{trigger-mode} \\
                      & \textbf{}  & from & to                                     & \textbf{observatory$^{b}$}    & \textbf{data} \\
\hline
\hline
\cite{2014ApJ...790..101S} & 1    &     2012-11-02   &     2012-11-02   &  AO            &        00537266000                                 \\
\cite{2016Natur.531..202S} &11    &     2012-11-02   &     2015-06-02   &  AO            &        00036376034                                 \\
\cite{2016ApJ...833..177S} & 6    &     2015-11-13   &     2015-12-08   &  AO, GBT       &        00051650045                                 \\
\cite{2018ApJ...863..150S} & 3    &     2016-08-20   &     2016-08-20   &  Effelsberg    &          n/a                                       \\
\cite{2017ApJ...850...76L} & 9    &     2016-08-23   &     2016-09-18   &  AO, VLA       &        00034292075, 00020700007                    \\
\cite{2019ApJ...877L..19G} & 42   &     2016-09-13   &     2016-09-14   &  AO            &        00034712006, 00081872001, 00020700002       \\
\cite{2017ApJ...846...80S} & 13   &     2016-09-16   &     2017-01-12   &  AO, GTB       &        00085966015, 00074948011, 00732188000       \\
\cite{2018Natur.553..182M} & 17   &     2016-12-25   &     2017-08-26   &  AO, GTB       &        00087278003                                 \\
\cite{2017MNRAS.472.2800H} & 13   &     2017-01-16   &     2017-02-19   &  Effelsberg    &        00050100039                                 \\
\cite{2018MNRAS.481.2479M} & 5    &     2017-02-15   &     2017-03-02   &  AO            &        00034921009                                 \\
\cite{2018ApJ...863....2G} & 21   &     2017-08-26   &     2017-08-26   &  GBT           &          n/a                                       \\
\cite{2018ApJ...866..149Z} & 93   &     2017-08-26   &     2017-08-26   &  GBT           &          n/a                                       \\
\cite{2019ApJ...882L..18J} & 1    &     2018-11-19   &     2018-11-19   &  CHIME         &          n/a                                       \\	 
\hline                                                                                     
\multicolumn{6}{l}{$^{a}$ Number of detected bursts from \frbc.} \\
\multicolumn{6}{l}{$^{b}$ Arecibo Observatory (AO); Canadian Hydrogen Intensity Mapping Experiment (CHIME); Effelsberg Radio Telescope;  } \\
\multicolumn{6}{l}{\quad Robert C. Byrd Green Bank Telescope (GBT); Very Large Array (VLA).} \\%
\end{tabular}
\end{table*}

All the fast radio bursts in \frbc that had been reported in the literature (since 2014, till August 2019) were investigated. The sample and the corresponding references are listed in Table~\ref{T:bursts}. We also referred to the paper of \cite{2019arXiv190103484L} where part of the sample was also collected.  Since the radio bursts in the repeating FRB have a timescale of about milliseconds in the radio band, to search for high-energy short-term bursts in association with those radio bursts either simultaneously or quasi-simultaneously, we have to make use of the event mode data in the \swift/BAT archive. These BAT triggered event mode data have a time-resolution of about 100 $\mu$s, and the corresponding time stamp of each photon detected by BAT was recorded.

We first aimed at a search for potential X-ray bursts occurred simultaneously with those 234 fast radio bursts in the sample list in Table~\ref{T:bursts}. Simultaneous BAT triggered event mode data (spread in 14 observations of total exposure about 64~ks) were found available to 46 radio bursts out of the sample. However, \frbc was not always in the field of view of BAT during these 14 observations, 
It turned out that there was only one radio burst covered with simultaneous BAT exposure in the observation 00085966015, so we further analyzed the dataset in more detail. 
Beyond the investigation of BAT event data covering the radio burst, we also searched for any X-ray bursts occurred before or after the radio bursts in the event data of those 14 observations when BAT's field-of-view covered \frbc. This corresponds to a search for hard X-ray bursts when the repeating FRB was in active radio bursting phases.

\begin{figure*}
  \centering
  \subfigure[BAT high time resolution light curve simultaneous to the radio burst from \frbc binned in milliseconds.] {

    \includegraphics[width=16.0cm,angle=0]{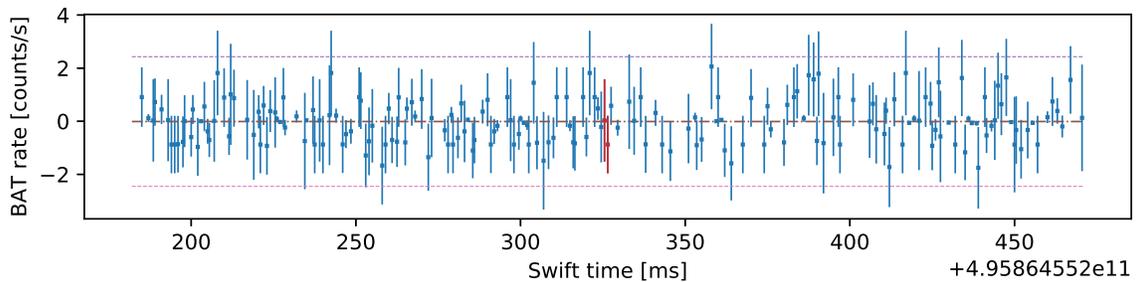}}    

\caption{The BAT light curve of \frbc with a time resolution of millisecond around the radio burst which occurred at 04:10:17.434 on 2016-09-18 (referenced to infinite frequency at the solar system barycenter) as reported by \cite{2017ApJ...846...80S}. This barycentric time is converted to the \swiftc time, the $\pm 1$ ms interval of which is marked in red in the light curve. Dashed lines denote the average and $\pm 3\sigma$ of the rates. See the corresponding BAT sky image in Fig.~\ref{momentMAP}.}
   \label{momentLC}
\end{figure*}

The event mode data were processed by the software package \textsc{HEASOFT} v6.19 following \emph{The SWIFT BAT Software Guide}.  The \textsc{batdetmask} tasks were used to produce the detector quality map from CALDB. The \textsc{batdetmask} tasks calculate the mask weighting for each event file. The \textsc{batbinevt} tasks were run again to produced the light curves (LCs) from event files corresponding to the specified direction of \frbc ~(R.A. 82$^\circ.9946$, decl. 33$^\circ.1479$) in the following four different energy bands, namely, 15--30, 30--60, 60--150, and 15--150~keV.  To generate the sky maps, the \textsc{batbinevt} tasks were run to convert those event lists to detector plane images (DPIs), and the \textsc{batfftimage} task was used to covert them into sky maps with photon counts or significance.  As a result, the total exposure time of the event mode data toward \frbc was about 2.1 ~ks. The FRBs have millisecond timescales in the radio band, and it is natural to search their X-ray counterparts on ms timescales, but not limited in ms timescale alone.  Our search for potential short-term bursts was conducted on four representing timescales, namely 1 ms, 10 ms, 100 ms, and 1000 ms. We decide our shortest timescale is 1 ms because small binning results in plenty of LC/image data and large uncertainty in flux, and hence demands more computing resource.

\begin{figure}
  \centering
  \subfigure[Simultaneous BAT sky image.]{\includegraphics[width=7.8cm,angle=0]{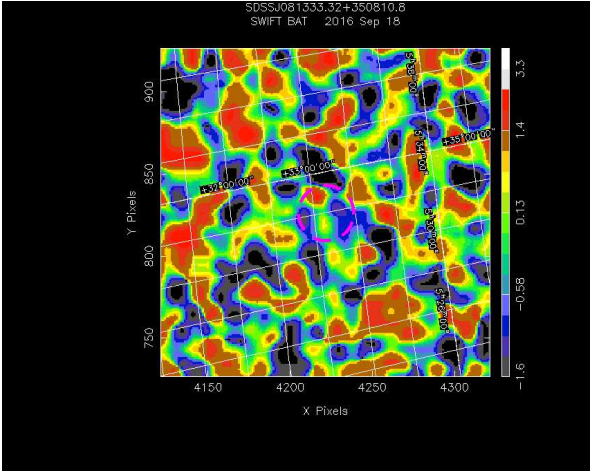}}
\caption{The BAT sky image in the direction of \frbc right at the moment when the radio bust occurred at 04:10:17.434 on 2016-09-18 (referenced to infinite frequency at the solar system barycenter) as reported by \cite{2017ApJ...846...80S}. The magenta dashed circle marks the nominal size of the point spread function of BAT coded-mask imaging. The corresponding BAT light curve is shown in Fig.~\ref{momentLC}.
}
   \label{momentMAP}
\end{figure}

In Figs.~\ref{momentLC} and \ref{momentMAP}, we present the BAT LC (15--150~keV; time bin of millisecond) and sky image which are simultaneous with the radio burst occurring on 2016-09-18 04:10:17.434 (referenced to infinite frequency at the solar system barycenter) from \frbc (reported by \cite{2017ApJ...846...80S}; see Table~\ref{T:bursts}). We convert times between the satellite and the barycenter with \textsc{barycorr} in \textsc{FTOOLS}, taking into account the relative locations of the satellite, the geocenter, and the barycenter. The LC time marked in red in Fig.~\ref{momentLC} is the result of converting the FRB barycentric time to that at \swiftc satellite. The significances of the two BAT rate measurements in the $\pm 1$ ms time window of the radio burst are both below 3$\sigma$ as shown in the LC (see in Fig.~\ref{momentLC} the dashed lines for the average and $\pm 3\sigma$ of the rates, and red markers for the $\pm 1$ ms time window) and the sky image (see in Fig.~\ref{momentMAP} the red color indicating significance below 3$\sigma$, and the magenta dashed circle marking the location of \frb). Below we put constraints on the X-ray flux simultaneous with the radio burst. Conservatively, the measurement with a larger rate value and a larger uncertainty is used for estimating an upper-limit of flux. The 3$\sigma$ upper limit of the count rate (= 3 $\times$ standard deviation of the rates; binned in milliseconds) we derived is 4.6 cts/s, which is equivalent to $6.9 \times 10^{-7}$erg~cm$^{-2}$s$^{-1}$ in the entire energy band of 15--150 keV if we assume an energy spectrum with a photoindex of 2. The count rates in the 10 ms, 100 ms, 1000 ms binned LCs at the radio burst time are checked, too, All of them below 3$\sigma$ of the rates, which correspond to 1.2, 0.38, and 0.091 cts/s.

\begin{figure*}
\centering
\subfigure[burst 2 LC, time bin of 1 ms.]{
\includegraphics[width=16cm]{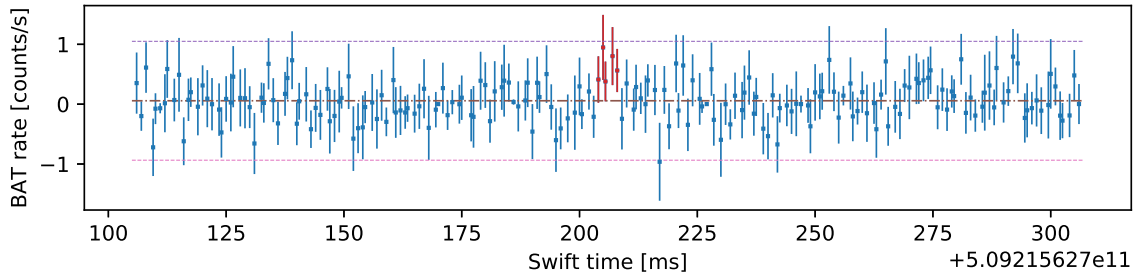}\label{pulseLCms}}

\caption{The BAT \frbc LC before and after the X-ray pulse (in red markers) with the highest s/n ratio among the search results as reported in Table~\ref{T:Xbursts}. Dashed lines denote the average and $\pm 3\sigma$ of the rates. See the BAT sky image of this pulse in Fig.~\ref{pulseMAP}.}
\label{pulseLC}
\end{figure*}

For the processing of the event mode data, we refer to the methods described in paragraph 4, section 2 in our previous paper of \cite{2019ApJ...885...55S}. To avoid complexities due to bright sources entering into the BAT field-of-view,  the following segments were excluded from the original BAT event data in our searches based on We are able to put an upper limitour investigations: 

(1) large dips occasionally appearing in the LCs with an interval within 1 time bin and an amplitude larger than the standard deviation. After applying these two conditions, about 85\% of the data remains, and 

(2) those segments with low signal-to-noise ratios: \\
$ | $ count rate / statistical error $| < 1$. \\

Using these criteria, we obtained the actual segments of the event data for our burst search. We had performed searches for potential X-ray bursts in the data sets on times scales of 1 ms, 10 ms, 100 ms and 1000 ms, as described in paragraph 5, section 2 of the previous paper \cite{2019ApJ...885...55S}.

After the computation of fluences ($= \displaystyle \sum_i^{\rm pulse} {\rm rate}_i$; called \emph{pulse integrals}, PIs) of candidate bursts, we then checked those bursts with fluences above a threshold. In the current study, we set the fluence threshold as: \\
0.0035 counts = $0.1\times$ crab rate $\times$ 1 sec. 
For setting up this criteria, we refer to the empirical rules used in the \swift/BAT project of transient monitoring\footnote{\url{https://swift.gsfc.nasa.gov/results/transients/}}, where sources are considered detected under the following circumstances: (a) the mean rate is at least 0.003 crab rate, or (b) the peak rate (1-day binned) for the source is at least 0.03 crab rate with at least a 7-sigma significance. When we applied 0.03 crab to our short-duration (1s--1ms) pulse search, there are too many noisy pulses from background fluctuation, so that we need to enhance the threshold. Furthermore, we check the s/n ratio ( $= \displaystyle \sum_i^{\rm pulse} {\rm rate}_i/\sqrt{\sum_i^{\rm pulse} {\rm error}_{\rm stst., i}^2}$ ): threshold of s/n ratio 5.0. Whenever there is a candidate burst above these two thresholds, a sky map corresponding to the specific time range is then generated for checking whether the candidate burst has an astrophysical origin. In some cases, certain instrumental effects caused significant flux fluctuation at the edge of the sky map (due to small-portion illumination of the detectors by the sources at the edge of the field of view). In some cases, cosmic ray events probably caused widespread illumination of the detectors. We exclude these events and remove the corresponding time intervals in the LCs.


\section{Results} 

\begin{table}
\caption{X-ray pulses with the highest s/n ratios from the search.}\label{T:Xbursts}
\centering
\begin{tabular}{ccccc}
\hline 
  & \multicolumn{2}{c}{pulse time$^{a}$}   & pulse integral$^{b}$   & pulse   \\
  & start time (s) & stop time (s)         &  (cts)                 & s/n ratio$^{c}$ \\
\hline
\hline
 \\
\multicolumn{5}{c}{search in 1 s binned and 0.1 s binned LCs} \\
\hline
\multicolumn{5}{c}{no candidate found} \\
\\
\multicolumn{5}{c}{search in 0.01 s binned LCs} \\
\hline
1 & 16539.365 &	 16539.405$^{f}$ 	& 0.00762 &	 	 3.63 	  \\
\\
\multicolumn{5}{c}{search in 0.001 s binned LCs}  \\
\hline
2 & 15627.204 &	 15627.209$^{f}$ 	& 0.00500 &	 	 3.77	  \\	 
3 & 16167.264 &	 16167.269$^{f}$ 	& 0.00503 &	 	 3.70 	  \\	 
4 & 16567.854 &	 16567.859$^{f}$ 	& 0.00449 &	 	 3.69 	  \\	 
5 & 22030.669 &	 22030.674$^{f}$ 	& 0.00470 &	 	 3.73 	  \\	 
\hline                                                              \multicolumn{5}{l}{$^{a}$ \swiftc time in seconds - 509200000} \\                       
\multicolumn{5}{l}{$^{b}$ $= \displaystyle \sum_i^{\rm pulse} {\rm rate}_i$ ; \quad  \quad $^{c}$ $= \displaystyle \sum_i^{\rm pulse} {\rm rate}_i/\sqrt{\sum_i^{\rm pulse} {\rm error}_{\rm stst., i}^2}$}   \\
\multicolumn{5}{l}{The pulse searching was conducted with the following conditions: } \\%
\multicolumn{5}{l}{\quad  threshold of pulse integral 0.0035 = $0.1\times$ crab unit $\times$ 1 s;} \\%
\multicolumn{5}{l}{\quad  threshold of s/n ratio 3.5.} \\%
\multicolumn{5}{l}{$^{f}$ In observation 00050100039 (obs. id).} \\
\end{tabular}
\end{table}

\begin{figure}
\centering
\subfigure[X-ray pulse 2, ms timescale]{
\includegraphics[width=7.8cm]{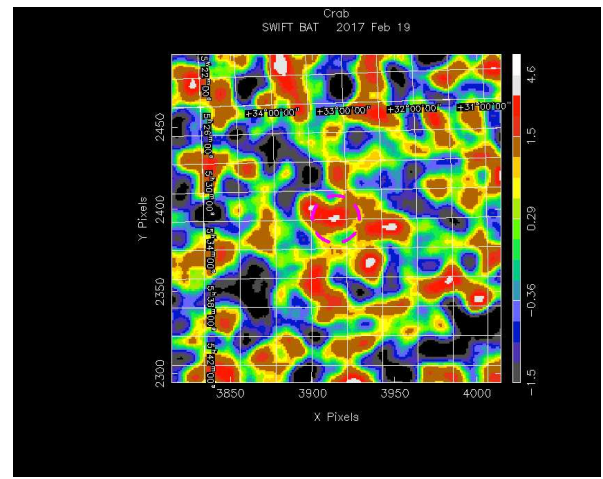}\label{pulseMAPms}}

\caption{The BAT sky image in the direction of \frbc taken during the time periods of the X-ray pulse with the highest S/N ratio among the searches as reported in Table~\ref{T:Xbursts}. The magenta dashed circle marks the nominal size of the point spread function of BAT coded-mask imaging. See the BAT LC before, on, and after the moment in Fig.~\ref{pulseLC}. }
\label{pulseMAP}
\end{figure}

As a result of searching in 1 ms, 10 ms, 100 ms and 1000 ms timescales with the threshold of s/n ratio 5.0, there is no candidate passing all the filters. To be careful with our search, we select pulses with highest s/n ratios to check what they look like. As we lower the threshold of s/n ratio to 3.5, five pulses are left, as reported in Table~\ref{T:Xbursts}. There is one found from the search in the 10 ms timescale; four from that in the 1 ms timescale. All of them have the s/n ratios between 3.5 and 4, in observation 00050100039, which might be associated with the radio burst occurring at 2017-02-19 16:37:48.114137 (referenced to infinite frequency at the solar system barycenter) reported by \cite{2017MNRAS.472.2800H}. All the pulses are found in this observation because the condition of BAT exposing toward \frbc in this observation was the best among all. Amongst the five pulses, pulse 2 has the shortest separation of about 800~s after the radio burst. The other pulses have separations of about 930~s--7200~s. They all happen to be after the the radio burst because the radio burst lied in the beginning of the BAT observation. We present the LCs and sky images for pulse 2 (1 ms timescale) in Figs.~\ref{pulseLC} and \ref{pulseMAP}. In the skymap of the X-ray pulse, we see that in the source direction the significance indicated by colors is only slightly higher than that in other regions, and cannot regard them as astrophysical bursts.



For evaluation of the statistical significance, we calculate the number of false positives above a given threshold to be expected from screening those BAT count rates below null by assuming that the count rates are affected by purely uncorrelated Gaussian noise. The number of false positive pulses expected above the s/n ratio of 3.5 for the 10 ms timescale is 0.89, and for the 1 ms timescale 7.6. The number of candidates from the search is 1 for the 10 ms timescale, and 4 for the 1 ms timescale. We find that the candidates are compatible with being statistical fluctuations.

\section{Conclusions} \label{dc}

We have searched for simultaneous short duration hard X-ray bursts as well as quai-simultaneous short duration bursts in the direction of \frbc in the \swift/BAT archival event mode data in which radio bursts were detected from \frb. 
There were existing BAT X-ray event mode data taken simultaneously to a radio burst detected in \frb, but there was no X-ray signal over $3\sigma$ found in the event of the radio burst. We are able to put an upper limit of $6.9 \times 10^{-7}$erg~cm$^{-2}$s$^{-1}$ of a hard X-ray burst in the energy band 15--150~keV simultaneous to the radio burst from the repeating \frb, assuming a photo index of 2. This limit on flux, converted into luminosity assuming isotropy and given the distance to the source \citep[$\sim$970\,Mpc][]{2017ApJ...834L...7T}, is $7.8 \times 10^{49}$erg~s$^{-1}$. In conclusion, we have not found evidence that supports that the source of FRB~121102 radiates strong hard X-ray burst emission detectable with current X-ray instrument like Swift/BAT simultaneously with those radio bursts.

 We noticed that recent detection of a bright radio burst (with two pulses) from the magnetar SGR 1935+2154 \citep{2020ATel13681....1S, 2020ATel13684....1B} while it was in active bursting phase \citep[including \swift/BAT;][]{2020ATel13675....1P} and subsequent identifications of its association with X-ray and soft gamma-ray  bursts \citep{2020ATel13685....1M, 2020ATel13686....1T, 2020ATel13687....1Z} have confirmed that X-ray bursts or burst activities could be related to some FRBs in the local distances. The measured count rate with BAT from SGR 1935+2154 was 350k counts/s on a 1 second timescale over the full detector sensitivity range. If we consider to put it as far away as \frbc \citep[SGR 1935+2154: $<10$ kpc; \frb: $\sim$970\,Mpc][]{2016MNRAS.460.2008K,2017ApJ...834L...7T}, then the predicted flux is still below the upper limit that we derived in this paper.   

\acknowledgments \label{Acknowledgments}

We note that the topic is very relevant to the recent breakthrough made through the study of SGR 1935+2154, but still any X-ray bursts in possible association with the FRBs of the same mechanism but originated at cosmological distances would still be well below the sensitivity of current wide field-of-view X-ray instrument like Swift/BAT.  W.Y. would like to thank Josh Grindlay of Harvard University and Craig Markwardt of GSFC and Jean in't Zand of SRON for stimulating discussions in the past on \swiftc/BAT data. W.Y. and S.Y.S. would like to thank David Palmer of LANL for very valuable comments on our previous analysis of BAT data. This work was supported in part by the National Program on Key Research and Development Project (Grant No. 2016YFA0400804) and the National Natural Science Foundation of China (grant Nos. 11333005, U1838203, and 11822302). S.Y.S. thank Zhen Yan and Stefano Rapisarda for helpful discussions for this study. S.Y.S. also acknowledges support from the International Postdoctoral Exchange Fellowship Recruitment Program of China Postdoctoral Council in 2018. W.Y. also acknowledges the support by the FAST Scholar fellowship in 2017, which is supported by special funding for advanced users, budgeted and administrated by Center for Astronomical Mega-Science, Chinese Academy of Sciences (CAMS). In this work, the authors used the data supplied by NASA's High Energy Astrophysics Science Archive Research Center in the US. The authors used the data products from the \swiftc mission, which is funded by NASA.


\bibliography{FRB121102X}{}

\begin{thebibliography}{}
\expandafter\ifx\csname natexlab\endcsname\relax\def\natexlab#1{#1}\fi
\providecommand{\url}[1]{\href{#1}{#1}}
\providecommand{\dodoi}[1]{doi:~\href{http://doi.org/#1}{\nolinkurl{#1}}}
\providecommand{\doeprint}[1]{\href{http://ascl.net/#1}{\nolinkurl{http://ascl.net/#1}}}
\providecommand{\doarXiv}[1]{\href{https://arxiv.org/abs/#1}{\nolinkurl{https://arxiv.org/abs/#1}}}

\bibitem[{{Bannister} {et~al.}(2019){Bannister}, {Deller}, {Phillips},
  {Macquart}, {Prochaska}, {Tejos}, {Ryder}, {Sadler}, {Shannon}, {Simha},
  {Day}, {McQuinn}, {North-Hickey}, {Bhandari}, {Arcus}, {Bennert}, {Burchett},
  {Bouwhuis}, {Dodson}, {Ekers}, {Farah}, {Flynn}, {James}, {Kerr}, {Lenc},
  {Mahony}, {O'Meara}, {Os{\l}owski}, {Qiu}, {Treu}, {U}, {Bateman},
  {Bock}, {Bolton}, {Brown}, {Bunton}, {Chippendale}, {Cooray}, {Cornwell},
  {Gupta}, {Hayman}, {Kesteven}, {Koribalski}, {MacLeod}, {McClure-Griffiths},
  {Neuhold}, {Norris}, {Pilawa}, {Qiao}, {Reynolds}, {Roxby}, {Shimwell},
  {Voronkov}, \& {Wilson}}]{2019Sci...365..565B}
{Bannister}, K.~W., {Deller}, A.~T., {Phillips}, C., {et~al.} 2019, Science,
  365, 565, \dodoi{10.1126/science.aaw5903}

\bibitem[{{Bochenek} {et~al.}(2020){Bochenek}, {Kulkarni}, {Ravi}, {McKenna},
  {Hallinan}, \& {Belov}}]{2020ATel13684....1B}
{Bochenek}, C., {Kulkarni}, S., {Ravi}, V., {et~al.} 2020, The Astronomer's
  Telegram, 13684, 1

\bibitem[{{Canal} {et~al.}(1990){Canal}, {Garcia}, {Isern}, \&
  {Labay}}]{1990ApJ...356L..51C}
{Canal}, R., {Garcia}, D., {Isern}, J., \& {Labay}, J. 1990, \apjl, 356, L51,
  \dodoi{10.1086/185748}

\bibitem[{{Cao} {et~al.}(2017){Cao}, {Yu}, \& {Dai}}]{2017ApJ...839L..20C}
{Cao}, X.-F., {Yu}, Y.-W., \& {Dai}, Z.-G. 2017, \apjl, 839, L20,
  \dodoi{10.3847/2041-8213/aa6af2}

\bibitem[{{Cao} {et~al.}(2018){Cao}, {Yu}, \& {Zhou}}]{2018ApJ...858...89C}
{Cao}, X.-F., {Yu}, Y.-W., \& {Zhou}, X. 2018, \apj, 858, 89,
  \dodoi{10.3847/1538-4357/aabadd}

\bibitem[{{Champion} {et~al.}(2016){Champion}, {Petroff}, {Kramer}, {Keith},
  {Bailes}, {Barr}, {Bates}, {Bhat}, {Burgay}, {Burke-Spolaor}, {Flynn},
  {Jameson}, {Johnston}, {Ng}, {Levin}, {Possenti}, {Stappers}, {van Straten},
  {Thornton}, {Tiburzi}, \& {Lyne}}]{2016MNRAS.460L..30C}
{Champion}, D.~J., {Petroff}, E., {Kramer}, M., {et~al.} 2016, \mnras, 460,
  L30, \dodoi{10.1093/mnrasl/slw069}

\bibitem[{{Chatterjee} {et~al.}(2017){Chatterjee}, {Law}, {Wharton},
  {Burke-Spolaor}, {Hessels}, {Bower}, {Cordes}, {Tendulkar}, {Bassa},
  {Demorest}, {Butler}, {Seymour}, {Scholz}, {Abruzzo}, {Bogdanov}, {Kaspi},
  {Keimpema}, {Lazio}, {Marcote}, {McLaughlin}, {Paragi}, {Ransom}, {Rupen},
  {Spitler}, \& {van Langevelde}}]{2017Natur.541...58C}
{Chatterjee}, S., {Law}, C.~J., {Wharton}, R.~S., {et~al.} 2017, \nat, 541, 58,
  \dodoi{10.1038/nature20797}

\bibitem[{{CHIME/FRB Collaboration} {et~al.}(2019{\natexlab{a}}){CHIME/FRB
  Collaboration}, {Amiri}, {Bandura}, {Bhardwaj}, {Boubel}, {Boyce}, {Boyle},
  {Brar}, {Burhanpurkar}, {Chawla}, {Cliche}, {Cubranic}, {Deng}, {Denman},
  {Dobbs}, {Fandino}, {Fonseca}, {Gaensler}, {Gilbert}, {Giri}, {Good},
  {Halpern}, {Hanna}, {Hill}, {Hinshaw}, {H{\"o}fer}, {Josephy}, {Kaspi},
  {Landecker}, {Lang}, {Masui}, {Mckinven}, {Mena-Parra}, {Merryfield},
  {Milutinovic}, {Moatti}, {Naidu}, {Newburgh}, {Ng}, {Patel}, {Pen},
  {Pinsonneault-Marotte}, {Pleunis}, {Rafiei-Ravandi}, {Ransom}, {Renard},
  {Scholz}, {Shaw}, {Siegel}, {Smith}, {Stairs}, {Tendulkar}, {Tretyakov},
  {Vanderlinde}, \& {Yadav}}]{2019Natur.566..230C}
{CHIME/FRB Collaboration}, {Amiri}, M., {Bandura}, K., {et~al.}
  2019{\natexlab{a}}, \nat, 566, 230, \dodoi{10.1038/s41586-018-0867-7}

\bibitem[{{CHIME/FRB Collaboration} {et~al.}(2019{\natexlab{b}}){CHIME/FRB
  Collaboration}, {Amiri}, {Bandura}, {Bhardwaj}, {Boubel}, {Boyce}, {Boyle},
  {.~Brar}, {Burhanpurkar}, {Cassanelli}, {Chawla}, {Cliche}, {Cubranic},
  {Deng}, {Denman}, {Dobbs}, {Fandino}, {Fonseca}, {Gaensler}, {Gilbert},
  {Gill}, {Giri}, {Good}, {Halpern}, {Hanna}, {Hill}, {Hinshaw}, {H{\"o}fer},
  {Josephy}, {Kaspi}, {Landecker}, {Lang}, {Lin}, {Masui}, {Mckinven},
  {Mena-Parra}, {Merryfield}, {Michilli}, {Milutinovic}, {Moatti}, {Naidu},
  {Newburgh}, {Ng}, {Patel}, {Pen}, {Pinsonneault-Marotte}, {Pleunis},
  {Rafiei-Ravandi}, {Rahman}, {Ransom}, {Renard}, {Scholz}, {Shaw}, {Siegel},
  {Smith}, {Stairs}, {Tendulkar}, {Tretyakov}, {Vanderlinde}, \&
  {Yadav}}]{2019Natur.566..235C}
---. 2019{\natexlab{b}}, \nat, 566, 235, \dodoi{10.1038/s41586-018-0864-x}

\bibitem[{{Connor} {et~al.}(2016){Connor}, {Sievers}, \&
  {Pen}}]{2016MNRAS.458L..19C}
{Connor}, L., {Sievers}, J., \& {Pen}, U.-L. 2016, \mnras, 458, L19,
  \dodoi{10.1093/mnrasl/slv124}

\bibitem[{{Dai} \& {Lu}(1998)}]{1998PhRvL..81.4301D}
{Dai}, Z.~G., \& {Lu}, T. 1998, \prl, 81, 4301,
  \dodoi{10.1103/PhysRevLett.81.4301}

\bibitem[{{Dai} {et~al.}(2017){Dai}, {Wang}, \& {Yu}}]{2017ApJ...838L...7D}
{Dai}, Z.~G., {Wang}, J.~S., \& {Yu}, Y.~W. 2017, \apjl, 838, L7,
  \dodoi{10.3847/2041-8213/aa6745}

\bibitem[{{Dai} {et~al.}(2006){Dai}, {Wang}, {Wu}, \&
  {Zhang}}]{2006Sci...311.1127D}
{Dai}, Z.~G., {Wang}, X.~Y., {Wu}, X.~F., \& {Zhang}, B. 2006, Science, 311,
  1127, \dodoi{10.1126/science.1123606}

\bibitem[{{Eatough} {et~al.}(2013){Eatough}, {Falcke}, {Karuppusamy}, {Lee},
  {Champion}, {Keane}, {Desvignes}, {Schnitzeler}, {Spitler}, {Kramer},
  {Klein}, {Bassa}, {Bower}, {Brunthaler}, {Cognard}, {Deller}, {Demorest},
  {Freire}, {Kraus}, {Lyne}, {Noutsos}, {Stappers}, \&
  {Wex}}]{2013Natur.501..391E}
{Eatough}, R.~P., {Falcke}, H., {Karuppusamy}, R., {et~al.} 2013, \nat, 501,
  391, \dodoi{10.1038/nature12499}

\bibitem[{{Gajjar} {et~al.}(2018){Gajjar}, {Siemion}, {Price}, {Law},
  {Michilli}, {Hessels}, {Chatterjee}, {Archibald}, {Bower}, {Brinkman},
  {Burke-Spolaor}, {Cordes}, {Croft}, {Enriquez}, {Foster}, {Gizani},
  {Hellbourg}, {Isaacson}, {Kaspi}, {Lazio}, {Lebofsky}, {Lynch}, {MacMahon},
  {McLaughlin}, {Ransom}, {Scholz}, {Seymour}, {Spitler}, {Tendulkar},
  {Werthimer}, \& {Zhang}}]{2018ApJ...863....2G}
{Gajjar}, V., {Siemion}, A.~P.~V., {Price}, D.~C., {et~al.} 2018, \apj, 863, 2,
  \dodoi{10.3847/1538-4357/aad005}

\bibitem[{{Gourdji} {et~al.}(2019){Gourdji}, {Michilli}, {Spitler}, {Hessels},
  {Seymour}, {Cordes}, \& {Chatterjee}}]{2019ApJ...877L..19G}
{Gourdji}, K., {Michilli}, D., {Spitler}, L.~G., {et~al.} 2019, \apjl, 877,
  L19, \dodoi{10.3847/2041-8213/ab1f8a}

\bibitem[{{Hardy} {et~al.}(2017){Hardy}, {Dhillon}, {Spitler}, {Littlefair},
  {Ashley}, {De Cia}, {Green}, {Jaroenjittichai}, {Keane}, {Kerry}, {Kramer},
  {Malesani}, {Marsh}, {Parsons}, {Possenti}, {Rattanasoon}, \&
  {Sahman}}]{2017MNRAS.472.2800H}
{Hardy}, L.~K., {Dhillon}, V.~S., {Spitler}, L.~G., {et~al.} 2017, \mnras, 472,
  2800, \dodoi{10.1093/mnras/stx2153}

\bibitem[{{Hessels} {et~al.}(2019){Hessels}, {Spitler}, {Seymour}, {Cordes},
  {Michilli}, {Lynch}, {Gourdji}, {Archibald}, {Bassa}, {Bower}, {Chatterjee},
  {Connor}, {Crawford}, {Deneva}, {Gajjar}, {Kaspi}, {Keimpema}, {Law},
  {Marcote}, {McLaughlin}, {Paragi}, {Petroff}, {Ransom}, {Scholz}, {Stappers},
  \& {Tendulkar}}]{2019ApJ...876L..23H}
{Hessels}, J.~W.~T., {Spitler}, L.~G., {Seymour}, A.~D., {et~al.} 2019, \apjl,
  876, L23, \dodoi{10.3847/2041-8213/ab13ae}

\bibitem[{{Josephy} {et~al.}(2019){Josephy}, {Chawla}, {Fonseca}, {Ng},
  {Patel}, {Pleunis}, {Scholz}, {Andersen}, {Bandura}, {Bhardwaj}, {Boyce},
  {Boyle}, {Brar}, {Cubranic}, {Dobbs}, {Gaensler}, {Gill}, {Giri}, {Good},
  {Halpern}, {Hinshaw}, {Kaspi}, {Landecker}, {Lang}, {Lin}, {Masui},
  {Mckinven}, {Mena-Parra}, {Merryfield}, {Michilli}, {Milutinovic}, {Naidu},
  {Pen}, {Rafiei-Ravandi}, {Rahman}, {Ransom}, {Renard}, {Siegel}, {Smith},
  {Stairs}, {Tendulkar}, {Vanderlinde}, {Yadav}, \&
  {Zwaniga}}]{2019ApJ...882L..18J}
{Josephy}, A., {Chawla}, P., {Fonseca}, E., {et~al.} 2019, \apjl, 882, L18,
  \dodoi{10.3847/2041-8213/ab2c00}

\bibitem[{{Kasen} \& {Bildsten}(2010)}]{2010ApJ...717..245K}
{Kasen}, D., \& {Bildsten}, L. 2010, \apj, 717, 245,
  \dodoi{10.1088/0004-637X/717/1/245}

\bibitem[{{Kozlova} {et~al.}(2016){Kozlova}, {Israel}, {Svinkin}, {Frederiks},
  {Pal'shin}, {Tsvetkova}, {Hurley}, {Goldsten}, {Golovin}, {Mitrofanov}, \&
  {Zhang}}]{2016MNRAS.460.2008K}
{Kozlova}, A.~V., {Israel}, G.~L., {Svinkin}, D.~S., {et~al.} 2016, \mnras,
  460, 2008, \dodoi{10.1093/mnras/stw1109}

\bibitem[{{Krimm} {et~al.}(2013){Krimm}, {Holland}, {Corbet}, {Pearlman},
  {Romano}, {Kennea}, {Bloom}, {Barthelmy}, {Baumgartner}, {Cummings},
  {Gehrels}, {Lien}, {Markwardt}, {Palmer}, {Sakamoto}, {Stamatikos}, \&
  {Ukwatta}}]{2013ApJS..209...14K}
{Krimm}, H.~A., {Holland}, S.~T., {Corbet}, R.~H.~D., {et~al.} 2013, \apjs,
  209, 14, \dodoi{10.1088/0067-0049/209/1/14}

\bibitem[{{Kulkarni} {et~al.}(2014){Kulkarni}, {Ofek}, {Neill}, {Zheng}, \&
  {Juric}}]{2014ApJ...797...70K}
{Kulkarni}, S.~R., {Ofek}, E.~O., {Neill}, J.~D., {Zheng}, Z., \& {Juric}, M.
  2014, \apj, 797, 70, \dodoi{10.1088/0004-637X/797/1/70}

\bibitem[{{Law} {et~al.}(2017){Law}, {Abruzzo}, {Bassa}, {Bower},
  {Burke-Spolaor}, {Butler}, {Cantwell}, {Carey}, {Chatterjee}, {Cordes},
  {Demorest}, {Dowell}, {Fender}, {Gourdji}, {Grainge}, {Hessels}, {Hickish},
  {Kaspi}, {Lazio}, {McLaughlin}, {Michilli}, {Mooley}, {Perrott}, {Ransom},
  {Razavi-Ghods}, {Rupen}, {Scaife}, {Scott}, {Scholz}, {Seymour}, {Spitler},
  {Stovall}, {Tendulkar}, {Titterington}, {Wharton}, \&
  {Williams}}]{2017ApJ...850...76L}
{Law}, C.~J., {Abruzzo}, M.~W., {Bassa}, C.~G., {et~al.} 2017, \apj, 850, 76,
  \dodoi{10.3847/1538-4357/aa9700}

\bibitem[{{Li} {et~al.}(2019){Li}, {Li}, {Zhang}, {Geng}, {Song}, {Huang}, \&
  {Yang}}]{2019arXiv190103484L}
{Li}, B., {Li}, L.-B., {Zhang}, Z.-B., {et~al.} 2019, arXiv e-prints.
\newblock \doarXiv{1901.03484}

\bibitem[{{Lorimer} {et~al.}(2007){Lorimer}, {Bailes}, {McLaughlin},
  {Narkevic}, \& {Crawford}}]{2007Sci...318..777L}
{Lorimer}, D.~R., {Bailes}, M., {McLaughlin}, M.~A., {Narkevic}, D.~J., \&
  {Crawford}, F. 2007, Science, 318, 777, \dodoi{10.1126/science.1147532}

\bibitem[{{Lyubarsky}(2014)}]{2014MNRAS.442L...9L}
{Lyubarsky}, Y. 2014, \mnras, 442, L9, \dodoi{10.1093/mnrasl/slu046}

\bibitem[{{MAGIC Collaboration} {et~al.}(2018){MAGIC Collaboration}, {Acciari},
  {Ansoldi}, {Antonelli}, {Arbet Engels}, {Arcaro}, {Baack}, {Babi{\'c}}, {},
  {Banerjee}, {Bangale}, {Barres de Almeida}, {Barrio}, {Becerra Gonz{\'a}lez},
  {Bednarek}, {Bernardini}, {Berti}, {Besenrieder}, {Bhattacharyya},
  {Bigongiari}, {Biland}, {Blanch}, {Bonnoli}, {Carosi}, {Ceribella},
  {Chatterjee}, {Colak}, {Colin}, {Colombo}, {Contreras}, {Cortina}, {Covino},
  {Cumani}, {D'Elia}, {da Vela}, {Dazzi}, {de Angelis}, {de Lotto}, {Delfino},
  {Delgado}, {di Pierro}, {Dom{\'{\i}}nguez}, {Dominis Prester}, {Dorner},
  {Doro}, {Einecke}, {Elsaesser}, {Fallah Ramazani}, {Fattorini},
  {Fern{\'a}ndez-Barral}, {Ferrara}, {Fidalgo}, {Foffano}, {Fonseca}, {Font},
  {Fruck}, {Gallozzi}, {Garc{\'{\i}}a L{\'o}pez}, {Garczarczyk}, {Gaug},
  {Giammaria}, {Godinovi{\'c}}, {}, {Guberman}, {Hadasch}, {Hahn}, {Hassan},
  {Herrera}, {Hoang}, {Hrupec}, {Inoue}, {Ishio}, {Iwamura}, {Kubo}, {Kushida},
  {Kuve{\v z}di{\'c}}, {}, {Lamastra}, {Lelas}, {Leone}, {Lindfors},
  {Lombardi}, {Longo}, {L{\'o}pez}, {L{\'o}pez-Oramas}, {Maggio}, {Majumdar},
  {Makariev}, {Maneva}, {Manganaro}, {Mannheim}, {Maraschi}, {Mariotti},
  {Mart{\'{\i}}nez}, {Masuda}, {Mazin}, {Minev}, {Miranda}, {Mirzoyan},
  {Molina}, {Moralejo}, {Moreno}, {Moretti}, {Neustroev}, {Niedzwiecki},
  {Nievas Rosillo}, {Nigro}, {Nilsson}, {Ninci}, {Nishijima}, {Noda},
  {Nogu{\'e}s}, {Paiano}, {Palacio}, {Paneque}, {Paoletti}, {Paredes},
  {Pedaletti}, {Pe{\~n}il}, {Peresano}, {Persic}, {Prada Moroni}, {Prandini},
  {Puljak}, {Garcia}, {Rhode}, {Rib{\'o}}, {Rico}, {Righi}, {Rugliancich},
  {Saha}, {Saito}, {Satalecka}, {Schweizer}, {Sitarek}, {{\v S}nidari{\'c}},
  {}, {Sobczynska}, {Somero}, {Stamerra}, {Strzys}, {Suri{\'c}}, {},
  {Tavecchio}, {Temnikov}, {Terzi{\'c}}, {}, {Teshima}, {Torres-Alb{\`a}},
  {Tsujimoto}, {Vanzo}, {Vazquez Acosta}, {Vovk}, {Ward}, {Will}, {Zari{\'c}},
  {Marcote}, {Spitler}, {Hessels}, {Kashiyama}, {Murase}, {Bosch-Ramon},
  {Michilli}, \& {Seymour}}]{2018MNRAS.481.2479M}
{MAGIC Collaboration}, {Acciari}, V.~A., {Ansoldi}, S., {et~al.} 2018, \mnras,
  481, 2479, \dodoi{10.1093/mnras/sty2422}

\bibitem[{{Margalit} {et~al.}(2019){Margalit}, {Berger}, \&
  {Metzger}}]{2019ApJ...886..110M}
{Margalit}, B., {Berger}, E., \& {Metzger}, B.~D. 2019, \apj, 886, 110,
  \dodoi{10.3847/1538-4357/ab4c31}

\bibitem[{{Margalit} \& {Metzger}(2018)}]{2018ApJ...868L...4M}
{Margalit}, B., \& {Metzger}, B.~D. 2018, \apjl, 868, L4,
  \dodoi{10.3847/2041-8213/aaedad}

\bibitem[{{Mereghetti} {et~al.}(2020){Mereghetti}, {Savchenko}, {Gotz},
  {Rodriguez}, {Ducci}, {Ferrigno}, {Bozzo}, {Borkowski}, \&
  {Bazzano}}]{2020ATel13685....1M}
{Mereghetti}, S., {Savchenko}, V., {Gotz}, D., {et~al.} 2020, The Astronomer's
  Telegram, 13685, 1

\bibitem[{{Metzger} {et~al.}(2017){Metzger}, {Berger}, \&
  {Margalit}}]{2017ApJ...841...14M}
{Metzger}, B.~D., {Berger}, E., \& {Margalit}, B. 2017, \apj, 841, 14,
  \dodoi{10.3847/1538-4357/aa633d}

\bibitem[{{Metzger} {et~al.}(2019){Metzger}, {Margalit}, \&
  {Sironi}}]{2019MNRAS.485.4091M}
{Metzger}, B.~D., {Margalit}, B., \& {Sironi}, L. 2019, \mnras, 485, 4091,
  \dodoi{10.1093/mnras/stz700}

\bibitem[{{Metzger} {et~al.}(2008){Metzger}, {Quataert}, \&
  {Thompson}}]{2008MNRAS.385.1455M}
{Metzger}, B.~D., {Quataert}, E., \& {Thompson}, T.~A. 2008, \mnras, 385, 1455,
  \dodoi{10.1111/j.1365-2966.2008.12923.x}

\bibitem[{{Michilli} {et~al.}(2018){Michilli}, {Seymour}, {Hessels}, {Spitler},
  {Gajjar}, {Archibald}, {Bower}, {Chatterjee}, {Cordes}, {Gourdji}, {Heald},
  {Kaspi}, {Law}, {Sobey}, {Adams}, {Bassa}, {Bogdanov}, {Brinkman},
  {Demorest}, {Fernandez}, {Hellbourg}, {Lazio}, {Lynch}, {Maddox}, {Marcote},
  {McLaughlin}, {Paragi}, {Ransom}, {Scholz}, {Siemion}, {Tendulkar}, {van
  Rooy}, {Wharton}, \& {Whitlow}}]{2018Natur.553..182M}
{Michilli}, D., {Seymour}, A., {Hessels}, J.~W.~T., {et~al.} 2018, \nat, 553,
  182, \dodoi{10.1038/nature25149}

\bibitem[{{Nomoto} \& {Kondo}(1991)}]{1991ApJ...367L..19N}
{Nomoto}, K., \& {Kondo}, Y. 1991, \apjl, 367, L19, \dodoi{10.1086/185922}

\bibitem[{{Palmer}(2020)}]{2020ATel13675....1P}
{Palmer}, D.~M. 2020, The Astronomer's Telegram, 13675, 1

\bibitem[{{Petroff} {et~al.}(2015){Petroff}, {Johnston}, {Keane}, {van
  Straten}, {Bailes}, {Barr}, {Barsdell}, {Burke-Spolaor}, {Caleb}, {Champion},
  {Flynn}, {Jameson}, {Kramer}, {Ng}, {Possenti}, \&
  {Stappers}}]{2015MNRAS.454..457P}
{Petroff}, E., {Johnston}, S., {Keane}, E.~F., {et~al.} 2015, \mnras, 454, 457,
  \dodoi{10.1093/mnras/stv1953}

\bibitem[{{Popov} \& {Postnov}(2010)}]{2010vaoa.conf..129P}
{Popov}, S.~B., \& {Postnov}, K.~A. 2010, in Evolution of Cosmic Objects
  through their Physical Activity, ed. H.~A. {Harutyunian}, A.~M. {Mickaelian},
  \& Y.~{Terzian}, 129--132.
\newblock \doarXiv{0710.2006}

\bibitem[{{Ravi} {et~al.}(2019){Ravi}, {Catha}, {D'Addario}, {Djorgovski},
  {Hallinan}, {Hobbs}, {Kocz}, {Kulkarni}, {Shi}, {Vedantham}, {Weinreb}, \&
  {Woody}}]{2019Natur.572..352R}
{Ravi}, V., {Catha}, M., {D'Addario}, L., {et~al.} 2019, \nat, 572, 352,
  \dodoi{10.1038/s41586-019-1389-7}

\bibitem[{{Scholz} \& {Chime/Frb Collaboration}(2020)}]{2020ATel13681....1S}
{Scholz}, P., \& {Chime/Frb Collaboration}. 2020, The Astronomer's Telegram,
  13681, 1

\bibitem[{{Scholz} {et~al.}(2016){Scholz}, {Spitler}, {Hessels}, {Chatterjee},
  {Cordes}, {Kaspi}, {Wharton}, {Bassa}, {Bogdanov}, {Camilo}, {Crawford},
  {Deneva}, {van Leeuwen}, {Lynch}, {Madsen}, {McLaughlin}, {Mickaliger},
  {Parent}, {Patel}, {Ransom}, {Seymour}, {Stairs}, {Stappers}, \&
  {Tendulkar}}]{2016ApJ...833..177S}
{Scholz}, P., {Spitler}, L.~G., {Hessels}, J.~W.~T., {et~al.} 2016, \apj, 833,
  177, \dodoi{10.3847/1538-4357/833/2/177}

\bibitem[{{Scholz} {et~al.}(2017){Scholz}, {Bogdanov}, {Hessels}, {Lynch},
  {Spitler}, {Bassa}, {Bower}, {Burke-Spolaor}, {Butler}, {Chatterjee},
  {Cordes}, {Gourdji}, {Kaspi}, {Law}, {Marcote}, {McLaughlin}, {Michilli},
  {Paragi}, {Ransom}, {Seymour}, {Tendulkar}, \&
  {Wharton}}]{2017ApJ...846...80S}
{Scholz}, P., {Bogdanov}, S., {Hessels}, J.~W.~T., {et~al.} 2017, \apj, 846,
  80, \dodoi{10.3847/1538-4357/aa8456}

\bibitem[{{Spitler} {et~al.}(2014){Spitler}, {Cordes}, {Hessels}, {Lorimer},
  {McLaughlin}, {Chatterjee}, {Crawford}, {Deneva}, {Kaspi}, {Wharton},
  {Allen}, {Bogdanov}, {Brazier}, {Camilo}, {Freire}, {Jenet},
  {Karako-Argaman}, {Knispel}, {Lazarus}, {Lee}, {van Leeuwen}, {Lynch},
  {Ransom}, {Scholz}, {Siemens}, {Stairs}, {Stovall}, {Swiggum},
  {Venkataraman}, {Zhu}, {Aulbert}, \& {Fehrmann}}]{2014ApJ...790..101S}
{Spitler}, L.~G., {Cordes}, J.~M., {Hessels}, J.~W.~T., {et~al.} 2014, \apj,
  790, 101, \dodoi{10.1088/0004-637X/790/2/101}

\bibitem[{{Spitler} {et~al.}(2016){Spitler}, {Scholz}, {Hessels}, {Bogdanov},
  {Brazier}, {Camilo}, {Chatterjee}, {Cordes}, {Crawford}, {Deneva}, {Ferdman},
  {Freire}, {Kaspi}, {Lazarus}, {Lynch}, {Madsen}, {McLaughlin}, {Patel},
  {Ransom}, {Seymour}, {Stairs}, {Stappers}, {van Leeuwen}, \&
  {Zhu}}]{2016Natur.531..202S}
{Spitler}, L.~G., {Scholz}, P., {Hessels}, J.~W.~T., {et~al.} 2016, \nat, 531,
  202, \dodoi{10.1038/nature17168}

\bibitem[{{Spitler} {et~al.}(2018){Spitler}, {Herrmann}, {Bower}, {Chatterjee},
  {Cordes}, {Hessels}, {Kramer}, {Michilli}, {Scholz}, {Seymour}, \&
  {Siemion}}]{2018ApJ...863..150S}
{Spitler}, L.~G., {Herrmann}, W., {Bower}, G.~C., {et~al.} 2018, \apj, 863,
  150, \dodoi{10.3847/1538-4357/aad332}

\bibitem[{{Sun} {et~al.}(2019){Sun}, {Yu}, {Yu}, {Mao}, \&
  {Lin}}]{2019ApJ...885...55S}
{Sun}, S., {Yu}, W., {Yu}, Y., {Mao}, D., \& {Lin}, J. 2019, \apj, 885, 55,
  \dodoi{10.3847/1538-4357/ab4420}

\bibitem[{{Tavani} {et~al.}(2020){Tavani}, {Ursi}, {Verrecchia}, {Casentini},
  {Pittori}, {Pilia}, {Cardillo}, {Piano}, {Bulgarelli}, {Fioretti},
  {Parmiggiani}, {Lucarelli}, {Donnarumma}, {Vercellone}, {Gianotti},
  {Trifoglio}, {Giuliani}, {Mereghetti}, {Caraveo}, {Perotti}, {Chen}, {Argan},
  {Costa}, {Del Monte}, {Evangelista}, {Feroci}, {Lazzarotto}, {Lapshov},
  {Pacciani}, {Soffitta}, {Vittorini}, {Di Cocco}, {Fuschino}, {Galli},
  {Labanti}, {Marisaldi}, {Pellizzoni}, {Trois}, {Barbiellini}, {Vallazza},
  {Longo}, {Morselli}, {Picozza}, {Prest}, {Lipari}, {Zanello}, {Cattaneo},
  {Rappoldi}, {Ferrari}, {Paoletti}, {Antonelli}, {Giommi}, {Salotti},
  {Valentini}, \& {D'Amico}}]{2020ATel13686....1T}
{Tavani}, M., {Ursi}, A., {Verrecchia}, F., {et~al.} 2020, The Astronomer's
  Telegram, 13686, 1

\bibitem[{{Tendulkar} {et~al.}(2017){Tendulkar}, {Bassa}, {Cordes}, {Bower},
  {Law}, {Chatterjee}, {Adams}, {Bogdanov}, {Burke-Spolaor}, {Butler},
  {Demorest}, {Hessels}, {Kaspi}, {Lazio}, {Maddox}, {Marcote}, {McLaughlin},
  {Paragi}, {Ransom}, {Scholz}, {Seymour}, {Spitler}, {van Langevelde}, \&
  {Wharton}}]{2017ApJ...834L...7T}
{Tendulkar}, S.~P., {Bassa}, C.~G., {Cordes}, J.~M., {et~al.} 2017, \apjl, 834,
  L7, \dodoi{10.3847/2041-8213/834/2/L7}

\bibitem[{{The CHIME/FRB Collaboration} {et~al.}(2019){The CHIME/FRB
  Collaboration}, {:}, {Andersen}, {Bandura}, {Bhardwaj}, {Boubel}, {Boyce},
  {Boyle}, {Brar}, {Cassanelli}, {Chawla}, {Cubranic}, {Deng}, {Dobbs},
  {Fandino}, {Fonseca}, {Gaensler}, {Gilbert}, {Giri}, {Good}, {Halpern},
  {H{\"o}fer}, {Hill}, {Hinshaw}, {Josephy}, {Kaspi}, {Kothes}, {Landecker},
  {Lang}, {Li}, {Lin}, {Masui}, {Mena-Parra}, {Merryfield}, {Mckinven},
  {Michilli}, {Milutinovic}, {Naidu}, {Newburgh}, {Ng}, {Patel}, {Pen},
  {Pinsonneault-Marotte}, {Pleunis}, {Rafiei-Ravandi}, {Rahman}, {Ransom},
  {Renard}, {Scholz}, {Siegel}, {Singh}, {Smith}, {Stairs}, {Tendulkar},
  {Tretyakov}, {Vanderlinde}, {Yadav}, \& {Zwaniga}}]{2019arXiv190803507T}
{The CHIME/FRB Collaboration}, {:}, {Andersen}, B.~C., {et~al.} 2019, arXiv
  e-prints.
\newblock \doarXiv{1908.03507}

\bibitem[{{Thornton} {et~al.}(2013){Thornton}, {Stappers}, {Bailes},
  {Barsdell}, {Bates}, {Bhat}, {Burgay}, {Burke-Spolaor}, {Champion}, {Coster},
  {D'Amico}, {Jameson}, {Johnston}, {Keith}, {Kramer}, {Levin}, {Milia}, {Ng},
  {Possenti}, \& {van Straten}}]{2013Sci...341...53T}
{Thornton}, D., {Stappers}, B., {Bailes}, M., {et~al.} 2013, Science, 341, 53,
  \dodoi{10.1126/science.1236789}

\bibitem[{{Woosley}(2010)}]{2010ApJ...719L.204W}
{Woosley}, S.~E. 2010, \apjl, 719, L204, \dodoi{10.1088/2041-8205/719/2/L204}

\bibitem[{{Yu} {et~al.}(2018){Yu}, {Liu}, \& {Dai}}]{2018ApJ...861..114Y}
{Yu}, Y.-W., {Liu}, L.-D., \& {Dai}, Z.-G. 2018, \apj, 861, 114,
  \dodoi{10.3847/1538-4357/aac6e5}

\bibitem[{{Yu} {et~al.}(2017){Yu}, {Zhu}, {Li}, {L{\"u}}, \&
  {Zou}}]{2017ApJ...840...12Y}
{Yu}, Y.-W., {Zhu}, J.-P., {Li}, S.-Z., {L{\"u}}, H.-J., \& {Zou}, Y.-C. 2017,
  \apj, 840, 12, \dodoi{10.3847/1538-4357/aa6c27}

\bibitem[{{Zhang} \& {M{\'e}sz{\'a}ros}(2001)}]{2001ApJ...552L..35Z}
{Zhang}, B., \& {M{\'e}sz{\'a}ros}, P. 2001, \apjl, 552, L35,
  \dodoi{10.1086/320255}

\bibitem[{{Zhang} {et~al.}(2020){Zhang}, {Tuo}, {Xiong}, {Li}, {Xiao}, {Jia},
  {Li}, {Ge}, {Luo}, {Li}, {Cai}, {Tan}, {Xue}, {Lu}, {Song}, {Liu}, {Chen},
  {Cao}, {Xu}, {Li}, {Lin}, \& {Zhang}}]{2020ATel13687....1Z}
{Zhang}, S.~N., {Tuo}, Y.~L., {Xiong}, S.~L., {et~al.} 2020, The Astronomer's
  Telegram, 13687, 1

\bibitem[{{Zhang} {et~al.}(2018){Zhang}, {Gajjar}, {Foster}, {Siemion},
  {Cordes}, {Law}, \& {Wang}}]{2018ApJ...866..149Z}
{Zhang}, Y.~G., {Gajjar}, V., {Foster}, G., {et~al.} 2018, \apj, 866, 149,
  \dodoi{10.3847/1538-4357/aadf31}

\end{thebibliography}
\bibliographystyle{aasjournal}


\end{document}